\documentstyle[multicol,aps,prb,epsfig]{revtex}

\renewcommand{\narrowtext}{\begin{multicols}{2} \global\columnwidth20.5pc}
\renewcommand{\widetext}{\end{multicols} \global\columnwidth42.5pc}

\input{epsf.tex}

\def\top#1{\vskip #1\begin{picture}(290,80)(80,500)\thinlines \put(
65,500){\line( 1, 0){255}}\put(320,500){\line( 0, 1){
5}}\end{picture}}
\def\bottom#1{\vskip #1\begin{picture}(290,80)(80,500)\thinlines \put(
330,500){\line( 1, 0){255}}\put(330,500){\line( 0, -1){
5}}\end{picture}}

\begin{document} 
 
\draft 
 
\title{Probing spin-charge separation in tunnel-coupled parallel
quantum wires} 
 
\author{U. Z\"ulicke and M. Governale} 
 
\address{Institut f\"ur Theoretische Festk{\"o}rperphysik, 
Universit{\"a}t Karlsruhe, D-76128 Karlsruhe, Germany} 
 
\date{\today} 
 
\maketitle 
 
\begin{abstract}
Interactions in one-dimensional (1D) electron systems are expected
to cause a dynamical separation of electronic spin and charge
degrees of freedom. A promising system for experimental
observation of this non-Fermi-liquid effect consists of two
quantum wires coupled via tunneling through an extended uniform
barrier. Here we consider the minimal model of an interacting 1D
electron system exhibiting spin-charge separation and calculate
the differential tunneling conductance as well as the
density-density response function. Both quantities exhibit
distinct strong features arising from spin-charge separation.
Our analysis of these features within the minimal model neglects
interactions between electrons of opposite chirality and applies
therefore directly to chiral 1D electron systems realized, e.g.,
at the edge of integer quantum-Hall systems. Physical insight
gained from our results is useful for interpreting current
experiment in quantum wires as our main conclusions still apply
with nonchiral interactions present. In particular, we discuss the
effect of charging due to applied voltages, and the possibility to
observe spin-charge separation in a time-resolved experiment.
\end{abstract} 
 
\pacs{PACS number(s): 73.63.Nm, 73.40.Gk, 71.10.Pm} 
 
\narrowtext

\section{Introduction}

One-dimensional (1D) electron systems are one of the theoretically
best-studied examples where interactions lead to strong
correlations such that low-energy excitations cannot be described
using Landau's Fermi-liquid concept.\cite{voit:reprog:94} As soon
as electron-electron interactions are switched on, electron-like
quasiparticles cease to exist at low energies, and the elementary
excitations are phonon-like charge and spin-density fluctuations,
and topological zero modes. Physical quantities of such a {\em
Luttinger liquid\/}\cite{fdmh:jpc:81} are determined by the
velocities,\cite{velocaveat} $v_\rho$ and $v_\sigma$, of the
charge and spin-density phonons, as well as additional velocity
parameters characterizing the energy of topological modes. The
most striking non-Fermi-liquid feature is exhibited by the
single-electron spectral function ${\mathcal{A}}(k,\varepsilon)$
which is basically a measure of the integrity of an electron as
an elementary excitation in a many-body system.\cite{mahan} In the
absence of electron-electron interactions, the spectral function
is given by ${\mathcal{A}}^{(0)}(k,\varepsilon)=2\pi\delta(
\varepsilon-\varepsilon_k)$ where $\varepsilon_k$ is the
electronic band dispersion. For an interacting system, the
spectral function is generally broadened. However, in a Fermi
liquid, ${\mathcal{A}}(k,\varepsilon)$ still exhibits a distinct
single-electron-like peak, making it possible to represent the
system of interacting electrons as a system of non-interacting
quasiparticles that carry the same quantum numbers as free
electrons. Such a quasiparticle peak is absent in the spectral
function of a Luttinger liquid. Instead, a characteristic {\em
double-peak\/} structure appears.\cite{med:prb:92,voit:prb:93} The
existence of the two peaks whose energy dispersions follow those
of the elementary charge and spin-density excitations can be
interpreted as the dynamical break-up of the electron into two
independent entities representing its spin and charge degrees of
freedom.\cite{powerlaw}

Experimental verification of spin-charge separation essentially
requires a direct measurement of ${\mathcal{A}}(k,\varepsilon)$,
e.g., by photoemission\cite{photo} or
tunneling\cite{alt:prl:99,balents} spectroscopy. Recent progress
in fabrication techniques has made it possible to create a system
of two parallel quantum wires that are separated by a long and
clean tunnel barrier.\cite{amir} Uniformity of tunneling between
the two quantum wires, labeled U(pper) and L(ower), respectively,
implies that canonical momentum is approximately conserved in a
single tunneling event. The possibility to tune canonical versus
kinetic momentum by an external magnetic field makes it possible
to perform momentum-resolved tunneling studies.\cite{before} For
example, in a 1D Fermi liquid, resonances appear in the
magnetic-field dependence of the linear tunneling conductance
whenever Fermi points from different wires
overlap,\cite{uz:01b,llnores} i.e., when for any $\alpha,
\alpha^\prime=\pm$ the parameter 
\begin{equation}\label{resopara}
\kappa_{\alpha\alpha^\prime}=\frac{\pi}{2}\left(\alpha^\prime\,
n_{\text{U}}-\alpha\, n_{\text{L}}\right) + \frac{e}{\hbar}B\, d 
\end{equation}
vanishes. [Here, $n_{\text{U(L)}}$ denotes the 1D electron density
in the U (L) wire when no voltage is applied, $B$ is the magnetic
field applied perpendicular to the plane defined by the two wires,
and $d$ their separation.] Close to these resonance condition of
Eq.~(\ref{resopara}), the {\em differential\/} tunneling
conductance (DTC) as a function of voltage and magnetic field can
be expected to show features arising from spin-charge separation.
This motivates the first part of our study, detailed in
Sec.~\ref{pertsec}, where we calculate the DTC to lowest order in
perturbation theory close to the resonance point corresponding to
$\alpha=\alpha^\prime=+$ for a model 1D electron system where
interactions between electrons with opposite chirality are
neglected. It turns out that charging\cite{buttbefore} in response
to the applied voltage significantly alters tunneling
characteristics. We expect this finding to hold also in the
realistic case of non-chiral quantum wires.\cite{balentscomm}
Detailed expressions are given for the location of four
characteristic maxima in the DTC that are a manifestation of
spin-charge separation.\cite{addmaxcomm}

A perturbative treatment of tunneling is valid only for
calculating physical properties above an, in general,
interaction-dependent\cite{pertscale} energy scale. In the absence
of interactions, this scale is given by the tunneling strength 
$|t|$, and the regime where perturbation theory fails is
characterized by spatial and temporal oscillations in electronic
correlation functions.\cite{uz:01b} These oscillations result from
coherent electron motion between the two systems.\cite{dircoupl}
In Sec.~\ref{nonpert}, we provide a theoretical framework to treat
the nonperturbative regime {\em with interactions present\/} using
bosonization\cite{fink:prb:93,fabrizio} and
refermionization\cite{lin:prb:98,naudpapers} techniques. In
analogy with previous results,\cite{fabrizio} a characteristic
length scale $L_{\text{int}}=\pi\hbar|v_{\rho-}-v_\sigma|/|2 t|$
emerges from our calculation, measuring the relative strengths of
tunneling and interactions. On length scales shorter than
$L_{\text{int}}$, tunneling is irrelevant, i.e., it does not
affect the electronic structure of the two wires.\cite{fabrizio}
In the opposite limit of large length scales, however, we identify
spatial oscillations in the density response with a wave length
$L_t=\pi\hbar\sqrt{v_{\rho-} v_\sigma}/|t|$ that is renormalized
from its value $\pi\hbar v_{\text{F}}/|t|$ in the noninteracting
limit. In addition, we show a peculiar mode splitting to occur in
the density response that is similar to the one found
previously\cite{fink:prb:93,fabrizio} for the single-electron
Greens function. Besides characteristic charge-mode velocities
$v_{\rho\pm}$, an additional velocity $\bar v=2 v_{\rho-}
v_\sigma/(v_{\rho-}+v_\sigma)$ appears in the density response
that should be observable, in principle, in a time-resolved
experiment. Its existence is a manifestation of spin-charge
separation in the tunnel-coupled double-wire system. The naive
expectation that the
density response is sensitive just to the charge mode is satisfied
only in the previously mentioned limit where interactions
dominate tunneling.

In our study of spin-charge separation in tunnel-coupled quantum
wires, we consider a particular model for an interacting 1D
electron system. To be specific, the wires are assumed to be
parallel to the $x$ direction, located at $y=0$ and $z=
z_{\text{U,L}}$, respectively. The Hamiltonian is given by 
\begin{mathletters}\label{hamilton}
\begin{eqnarray}
H&=&H_0+H_{\text{tun}}+H_{\text{int}} \quad , \\
H_0 &=& \sum_{k\sigma\atop\alpha\beta} [\alpha\hbar
v_{\text{F}\beta}\left(k - \alpha k_{\text{F}\beta}\right) + 
\nu_\beta\, e V_\beta ] \, c_{k\sigma\alpha\beta}^\dagger c_{k
\sigma\alpha\beta}^{}\, ,\\
H_{\text{tun}} &=& t \sum_{k \sigma \alpha}\left\{
c_{k\sigma\alpha\text{U}}^\dagger c_{k\sigma\alpha\text{L}}^{}+
c_{k\sigma\alpha\text{L}}^\dagger c_{k\sigma\alpha\text{U}}^{}
\right\}\, , \label{htun} \\
H_{\text{int}} &=& \frac{1}{2 L} \sum_{\alpha q \atop \beta
\beta^\prime} U_{\beta \beta^\prime}\, \bigg(\sum_\sigma
\varrho_{q\sigma\alpha\beta}\bigg) \,\bigg(\sum_{\sigma^\prime}
\varrho_{-q\sigma^\prime\alpha\beta^\prime}\bigg)\, .
\end{eqnarray}
\end{mathletters}
Indices $\beta,\beta^\prime=\text{U,L}$ distinguish between the
upper and lower wire, and $\alpha=\pm$ between right-movers and
left-movers. Spin quantum numbers are denoted by $\sigma,
\sigma^\prime$. Each wire's Fermi wave vector $k_{\text{F}\beta}=
\frac{\pi}{2}n_\beta + \frac{e B}{\hbar}\, z_\beta$ contains the
effect of a magnetic field\cite{Zeeman} $\vec B=B\,\hat y$. A
voltage $V_{\text{U(L)}}$ is applied to the U (L) wire, and (in
general, voltage-dependent) parameters $\nu_\beta$ measure the
resulting shift of electron bands.\cite{weakt} The wires are
coupled by a tunneling matrix element $t$ chosen to be real.
Fourier transforms of the density of spin-$\sigma$ electrons from
the $\alpha$ branch in the $\beta$ wire are denoted by $\varrho_{q
\sigma\alpha\beta}$. Interactions included in $H_{\text{int}}$
are {\em chiral}, i.e., only electrons {\em from the same
branch\/} (right-moving or left-moving) within each wire and
between the two wires interact. This model applies directly to
interacting edge channels in quantum-Hall
bilayers\cite{naudcomment} when each layer is at filling factor 2
and Zeeman splitting is negligible. Interactions between
left-moving and right-moving electrons which are present in real
quantum wires are not accounted for in our model. It turns out,
however, that strong features arising from spin-charge separation
are accurately described already by the chiral
model.\cite{voit:prb:93,fabrizio,alt:prl:99} For example, the
redistribution of spectral weight due to non-chiral interactions
leads only to small additional structure in the
DTC.\cite{balents,addmaxcomm} Possibilities to go beyond the
chiral model are discussed in Sec.~\ref{nonpert}.

\section{Perturbative treatment}\label{pertsec}

Results presented in this section are obtained within lowest-order
perturbation theory in the tunneling Hamiltonian displayed in
Eq.~(\ref{htun}). We focus on magnetic fields and voltages close
to the resonance point where only right-movers tunnel. [See
Eq.~(\ref{resopara}) for $\alpha=\alpha^\prime=+$, and the inset
of Fig.~\ref{bshift}.] In the following two subsections, we
provide details of the calculations and give results for the
differential tunneling conductance (DTC), respectively.

\subsection{Formalism}\label{pertsub}

A standard calculation~\cite{mahan} to lowest order of
perturbation theory in $H_{\text{tun}}$ yields the expression
\begin{equation}\label{tunncurr}
I = 4 e \frac{|t|^2}{\hbar^2}\, L\, \, \Im\text{m}\left\{\int_x
\int_\tau e^{i\omega\tau}{\mathcal J}(x, \tau)\right\}_{i\omega\to
e V + i\delta}
\end{equation}
for the tunneling current. Here we denoted the barrier length by
$L$  and the electrochemical-potential difference across the
barrier by $V=V_{\text{U}}-V_{\text{L}}$. The Matsubara Greens
function
\begin{eqnarray}
&&{\mathcal J}(x,\tau)=\nonumber \\
&&-\left\langle T_\tau\psi^\dagger_{\sigma+
\text{U}}(x, \tau)\psi_{\sigma+\text{L}}(x, \tau)
\psi^\dagger_{\sigma+\text{L}}(0,0)\psi_{\sigma+\text{U}}(0,0)
\right\rangle
\end{eqnarray}
can be calculated straightforwardly\cite{balents} using
bosonization methods. Here we used the notation $\psi_{\sigma+
\beta}(x,\tau)=\sum_{k}e^{ik x} c_{k\sigma+\beta}(\tau)$. At zero
temperature, we find
\widetext
\top{-2.8cm}
\begin{equation}\label{greensfunct}
\left.{\mathcal J}(x,\tau)\right|_{T=0}=\frac{\exp\left\{i x\left(
\frac{e V_{\text{L}}}{\hbar v_{\text{FL}}}(1-\nu_{\text{L}})-\frac
{e V_{\text{U}}}{\hbar v_{\text{FU}}}(1-\nu_{\text{U}})-
\kappa_{++}\right)\right\}}{(2\pi)^2(v_{\sigma\text{L}}\tau -
i x)^{\frac{1}{2}}(v_{\sigma\text{U}}\tau - i x)^{\frac{1}{2}}
(v_{\rho+}\tau - i x)^{\frac{1}{2}+\theta_r}(v_{\rho-}\tau - i x
)^{\frac{1}{2}-\theta_r}}\quad .
\end{equation}
%\bottom{-2.7cm}
\narrowtext
\noindent
Spin-charge separation is manifested by the occurrence of four
algebraic singularities in the Greens function ${\mathcal J}(x,
\tau)$. $v_{\sigma\beta}=v_{\text{F}\beta}$ are the velocities of
spin-density excitations in the two wires, which are unaffected by
interactions. The charge-density eigenmodes in the double-wire
system have velocities
\begin{equation}
v_{\rho\pm} = \frac{v_{\rho\text{U}} +v_{\rho\text{L}}}{2}\pm
\frac{|U_{\text{UL}}|}{\pi\hbar}\sqrt{1+r^2}
\end{equation}
that differ from the charge-mode velocities $v_{\rho\beta}=
v_{\text{F}\beta}+U_{\beta\beta}/(\pi\hbar)$ of the two respective
wires due to inter-wire interactions. Here $r=\pi|v_{\rho\text{U}}
-v_{\rho\text{L}}|/|2U_{\text{UL}}|$. Another consequence of
inter-wire interactions is the finite exponent $\theta_r=1/(2\sqrt
{1+r^2})$. Note that, in the limit of strong interactions where
$r\to 0$, the singularity in ${\mathcal J}(x,\tau)$ associated
with velocity $v_{\rho-}$ disappears, and the singularity for
$v_{\rho+}$ changes into a pole. $\kappa_{\text{++}}$ is the
resonance parameter, defined in Eq.~(\ref{resopara}), that
measures the distance of the Fermi points for right-movers in the
two wires at zero applied voltage.

In real quasi-1D systems, voltage-induced shifts of electron
dispersion curves, denoted here by $\nu_{\text{U(L)}} e
V_{\text{U(L)}}$, depend strongly on sample details. It is for
that reason that we treat $\nu_{\text{U(L)}}$ as free parameters
when calculating the differential tunneling conductance. Our
general discussion of charging effects is intended to serve as a
useful guide to interpret experimental data. At the same time, we
would like to point out, however, that it is possible to derive
explicit expressions for the parameters $\nu_{\text{U(L)}}$ for
the model specified by Eqs.~(\ref{hamilton}). Application and
generalization of previous studies\cite{LLcharge} of charging in
Luttinger liquids to our double-wire model system yields
\begin{equation}\label{shiftpara}
\nu_{\text{U}} = 1-\frac{v_{\rho\text{L}}v_{\text{FU}}}{v_{\rho+}
v_{\rho-}} + \frac{(v_{\rho+} - v_{\rho-})v_{\text{FU}}}{2\sqrt
{1+r^2}\, v_{\rho+} v_{\rho-}}\,\frac{V_{\text{L}}}{V_{\text{U}}}
\quad, 
\end{equation}
and an analogous expression for $\nu_{\text{L}}$. Note that
charging is strongly affected by inter-wire interactions. In
particular, in the limit of strong interactions where no charging
would occur for separated wires, inter-wire interactions can drive
the double-wire system into a regime where charging is restored.

In the absence of inter-wire interactions, the expression for the
tunneling current given in Eq.~(\ref{tunncurr}) reduces to the
familiar form
\widetext
\top{-2.8cm}
\begin{equation}\label{current}
I = \frac{2 e}{\hbar}\frac{|t|^2 L}{(2\pi)^2}
\int_{e V_{\text{L}}}^{e V_{\text{U}}} d\varepsilon\,
\int_{-\infty}^{\infty} d q \,\, {\mathcal A}_{\text{L}}\left(
q - \frac{e V_{\text{L}}}{\hbar v_{\text{FL}}}(1 - \nu_{\text{L}})
+ \kappa_{++}, \varepsilon - e V_{\text{L}}\right) \,\,{\mathcal
A}_{\text{U}}\left(q - \frac{e V_{\text{U}}}{\hbar v_{\text{FU}}}
(1 - \nu_{\text{U}}),\varepsilon - e V_{\text{U}}\right) \quad .
\end{equation}
The functions ${\mathcal A}_\beta(q,\varepsilon)$ are known
exactly;\cite{voit:prb:93} they are the spectral functions of
chiral interacting 1D electron systems, containing right-movers
only, that are parameterized by the appropriate pair of spin and
charge velocities $v_{\sigma\beta}$, $v_{\rho\beta}$:
\begin{equation}
{\mathcal A}_{\beta}(q,\varepsilon)=\frac{\Theta(q)\Theta(
\varepsilon - v_{\sigma\beta} q)\Theta(v_{\rho\beta} q -
\varepsilon) + \Theta(-q)\Theta(\varepsilon - v_{\rho\beta} q)
\Theta(v_{\sigma\beta} q - \varepsilon) }{\pi\sqrt{|\epsilon -
v_{\sigma\beta} q||\epsilon - v_{\rho\beta} q|}}\quad .
\end{equation}
\bottom{-2.7cm}
\narrowtext
\noindent
Results shown in Figs.~\ref{bshift}--\ref{breal} were calculated
using Eq.~(\ref{current}) for the case $V_{\text{U}}=-V_{\text{L}}
=V/2$. To simplify the numerical calculation of the tunneling
current, we have applied Eq.~(\ref{current}) also to the case with
inter-wire interactions present, which is depicted in the inset of
Fig.~\ref{breal}. This approximation simply amounts to neglecting
the correction $\theta_r$ to exponents in the Greens function
${\mathcal J}(x,\tau)$ shown in Eq.~(\ref{greensfunct}). The
{\em location\/} of the four maxima in the DTC that are exhibited
by the chiral model,\cite{addmaxcomm} determined by the
singularities of ${\mathcal J}(x,\tau)$, will still be reproduced
adequately as long as $\theta_r<1/2$. The size of spectral weight
{\em inbetween\/} maxima of the DTC, however, will not be given
correctly by this approximation.

\subsection{Results}

\begin{figure}
\centerline{\includegraphics[width=7.5cm]{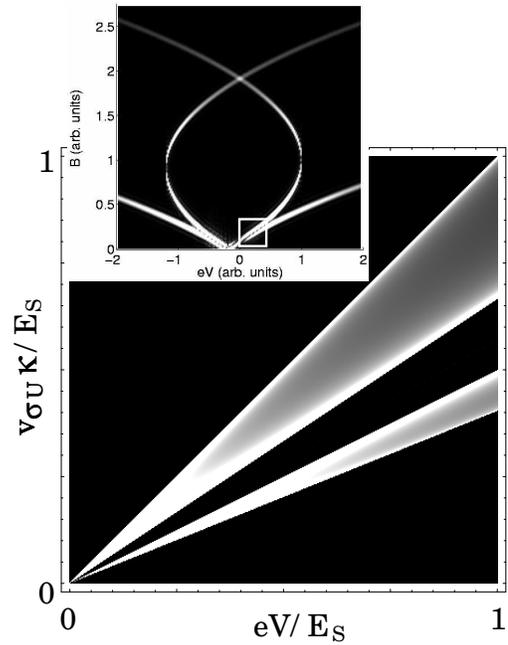}}
\vspace{0.3cm}
\caption{Density plot of the absolute value of the differential
tunneling conductance as function of voltage $V$ and magnetic
field in the absence of charging effects. The main figure is the
enlargement of the region indicated by a square in the inset. The
origin coincides with the resonance point $\kappa\equiv\kappa_{++}
=0$ [see Eq.~(\ref{resopara})] where Fermi points for right-movers
in the two wires overlap. Velocity parameters are chosen such that
$v_{\rho\text{U}}/v_{\sigma\text{U}}=1.5$, $v_{\sigma\text{L}}/
v_{\sigma\text{U}}=2$, $v_{\rho\text{L}}/v_{\sigma\text{U}}=2.5$,
and inter-wire interactions are neglected. The energy scale
$E_{\text{s}}$ is arbitrary. Slopes of strong (bright) features
are equal to the inverse of the velocity ratios given above.}
\label{bshift}
\end{figure}

\begin{figure}
\centerline{\includegraphics[width=7.5cm]{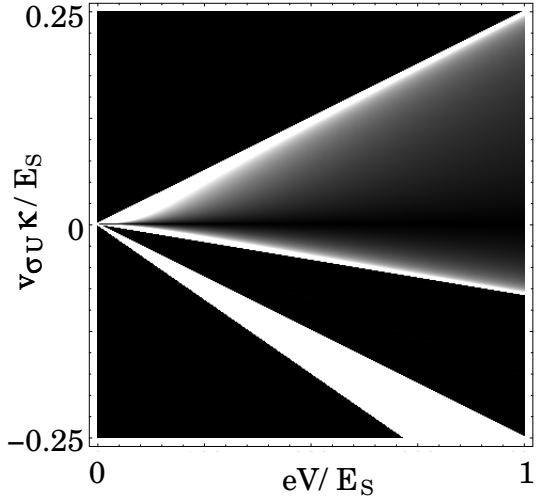}}
\vspace{0.3cm}
\caption{Density plot of the differential tunneling conductance
for the band-filling limit. Parameters are the same as in
Fig.~\ref{bshift}. Maxima follow the slopes $\pm\frac{1}{2}\left(1
-\frac{v_{\sigma\text{U}}}{v_{\sigma\text{L}}}\right)$, $\frac
{v_{\sigma\text{U}}}{v_{\rho\text{U}}}-\frac{1}{2}\left(1 + \frac
{v_{\sigma\text{U}}}{v_{\sigma\text{L}}}\right)$, and, $\frac
{v_{\sigma\text{U}}}{v_{\rho\text{L}}}-\frac{1}{2}\left(1 + \frac
{v_{\sigma\text{U}}}{v_{\sigma\text{L}}}\right)$, respectively.
Note that, for any choice of velocities, there is at least one
maximum following a negative slope.}
\label{bfill}
\end{figure}

\begin{figure}
\centerline{\includegraphics[width=7.5cm]{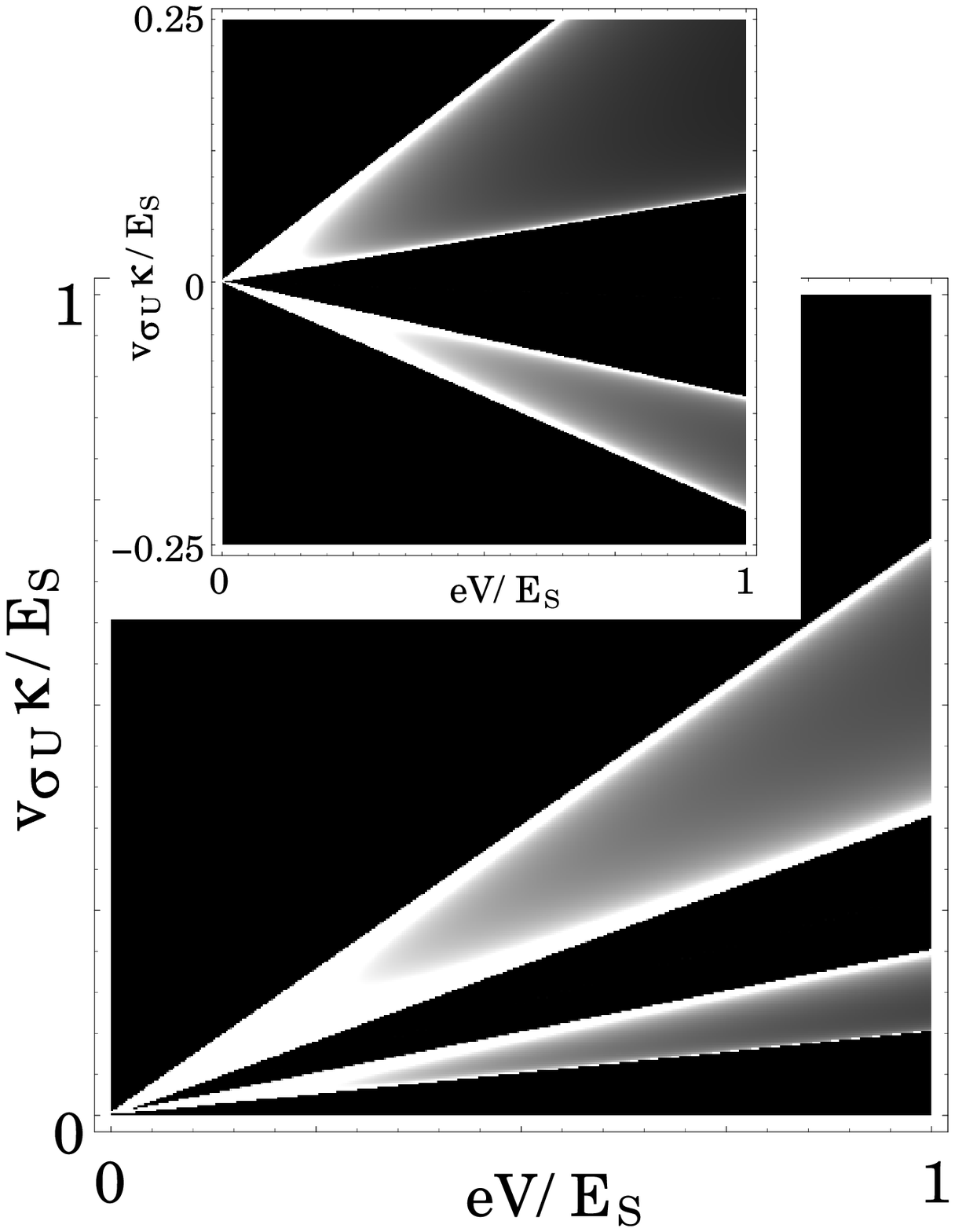}}
\vspace{0.3cm}
\caption{Density plot of the absolute value of the differential
tunneling conductance for an intermediate situation where the
applied voltage both shifts and fills electron bands. In the main
figure, we have chosen $\nu_{\text{U}}=\nu_{\text{L}} = 0.6$, in
addition to parameters used in Fig.~\ref{bshift}. Note that the
slopes of bright maxima are suppressed from their value in the
band-shifting case. The inset shows the result for the same system
with finite inter-wire interactions $U_{\text{UL}}/(\pi\hbar
v_{\sigma\text{U}})=0.25$ present and band-shifting parameters
$\nu_{\text{U}}=0.25$, $\nu_{\text{L}}=0.05$ calculated according
to Eq.~(\ref{shiftpara}).}
\label{breal}
\end{figure}

Strong features arise in the DTC from the algebraic singularities
of the Greens function shown in Eq.~(\ref{greensfunct}).
Spin-charge separation is manifested by maxima that form four
characteristic lines as a function of magnetic field and voltage.
This is seen, e.g., in Fig.~\ref{bshift} where we show the result
for the band-shifting limit\cite{uz:01b} where applied voltages
are assumed to shift electron bands without filling them:
$\nu_{\text{U}}=\nu_{\text{L}}=1$. The slopes of bright maxima are
given by the inverse of the charge and spin velocities. [The inset
of Fig.~\ref{bshift} shows the DTC  for noninteracting quantum
wires in the band-shifting limit,\cite{uz:01b} indicating the
region (inside the square box) that is enlarged in the main panel
where the effect of spin-charge separation can be observed.] In
the more general case, however, when the quantum wires are also
charged by applied voltages, the slope of these lines is changed.
We have analyzed the expression for the tunneling current for the
case of symmetric bias ($V_{\text{U}}=-V_{\text{L}}=V/2$). For an
analytic determination of characteristic equations for the lines
of maximal DTC as a function of magnetic field and voltage, we use
the simplistic replacement
$$ {\mathcal A}_{\beta}(q,\varepsilon)\to \delta(\varepsilon -
v_{\sigma\beta} q)+\delta(\varepsilon - v_{\rho\beta} q) \quad ,$$
obtaining
\begin{mathletters}\label{slopes}
\begin{eqnarray}
\frac{\kappa_{++}}{e V}&=& \frac{1+\nu_{\text{L}}}{2 v_{\sigma
\text{L}}} - \frac{1-\nu_{\text{U}}}{2 v_{\sigma\text{U}}} \quad,
\\ \frac{\kappa_{++}}{e V}&=& \frac{1}{v_{\rho\text{L}}} -
\frac{1-\nu_{\text{L}}}{2 v_{\sigma\text{L}}} - \frac{1-
\nu_{\text{U}}}{2 v_{\sigma\text{U}}}\quad ,
\end{eqnarray}
\end{mathletters}
and corresponding expressions where U and L are
exchanged.\cite{replacecav} We see that charging reduces the
characteristic slopes of maxima in the DTC, reaching their
smallest value in the band-filling limit ($\nu_{\text{U}}=
\nu_{\text{L}}=0$), which is shown in Fig~\ref{bfill}. As becomes
clear from specializing Eqs.~(\ref{slopes}) to the band-filling
limit, there is at least one negative slope for any quadruple of
velocities. This is very different from the band-shifting case
where the tunneling current vanishes in the region of negative
$\kappa_{++}$ and positive voltage for kinematic
reasons.\cite{alt:prl:99,balents}

For realistic quantum wires, an intermediate regime $0<
\nu_{\text{U}},\nu_{\text{L}}<1$ will be realized where applied
voltages both shift and fill electron bands. Examples are shown in
Fig~\ref{breal}. In  the main panel, the DTC for $\nu_{\text{U}}=
\nu_{\text{L}} = 0.6$ is displayed, illustrating the fact that
knowledge of charging properties is crucial for extracting the
charge and spin velocities from experimental data. The inset shows
the result to be expected for finite inter-wire interactions
(implying $v_{\rho+}/v_{\sigma\text{U}}=2.56$ and $v_{\rho-}/
v_{\sigma\text{U}}=1.44$), using $\nu_{\text{U(L)}}$ calculated
from Eq.~(\ref{shiftpara}).

\section{Beyond the perturbative regime}\label{nonpert}

Bosonization and refermionization\cite{vondelft} are powerful
methods enabling exact calculation of electronic correlation
functions for interacting 1D systems. Here we apply these to the
tunnel-coupled quantum-wire system described by the model
Hamiltonian~(\ref{hamilton}), extending previous
studies.\cite{fink:prb:93,fabrizio,lin:prb:98} In particular, we
give explicit expressions for the density response function, which
exhibits features due to spin-charge separation in the limit where
tunneling is relevant. We note that naive straighforward
calculation of the tunneling current within this model yields a
zero result,\cite{levitov} as perfect translational invariance
implies coherent motion of electrons between the wires and, hence,
vanishing current flow. Experimental detection of the tunneling
current requires leads to be attached to the system which breaks
translational invariance and results in a finite
current.\cite{uz:01b,balents} Perturbation theory actually
simulates this situation by excluding the possibility for
electrons to tunnel twice, which is adequate only if the tunneling
barrier is shorter than $\pi\hbar\sqrt{v_{\text{FU}}v_{\text{FL}}}
/|t|$.\cite{uz:01b}

\subsection{Reexpressing the Hamiltonian in new variables}

The Hamiltonian $H$ for the interacting double-wire system given
in Eqs.~(\ref{hamilton})  contains eight flavors of electrons,
distinguished by spin, wire index, and chirality. Following the
steps outlined in Appendix~\ref{bosref}, it is possible to rewrite
$H$ in terms of new degrees of freedom whose dynamics is simpler
than that of the original electrons:
\begin{mathletters}\label{rewrite}
\begin{eqnarray}\label{symmbos}
H&=&\sum_\alpha\Big\{\sum_{\gamma=\text{c,s}} H_\gamma^{(\alpha)}
+\sum_{j=1}^4 H_j^{(\alpha)}  + H_t^{(\alpha)}\nonumber \\ &&
\hspace{0.5cm}+H_\delta^{(\alpha)} + H_{\text{corr}}^{(\alpha)}
\Big\}\\
H_\gamma^{(\alpha)} &=& \frac{\hbar v_\gamma}{4\pi}\int d x\,\,
\left(\partial_x \Phi_\gamma^{(\alpha)}\right)^2 \quad , \\
H_j^{(\alpha)} &=& \frac{\hbar v_j}{2} \int d x \,\, \chi_j^{(
\alpha)}\, (-i\partial_x) \, \chi_j^{(\alpha)} \quad , \\
H_t^{(\alpha)} &=& 2 t \int d x \,\, i \chi_1^{(\alpha)}
\chi_2^{(\alpha)} \quad , \\
H_\delta^{(\alpha)} &=& \delta \int d x \,\, i \chi_2^{(\alpha)}
\chi_3^{(\alpha)} \quad , \\
H_{\text{corr}}^{(\alpha)} &=& \int d x \,\, \left[
\Delta_{\text{c}} \left(\partial_x \Phi_{\text{c}}^{(\alpha)}
\right)\, i \chi_3^{(\alpha)}\chi_2^{(\alpha)}\right.\nonumber \\
&& \hspace{0.5cm} \left. + \Delta_{\text{s}} \left(\partial_x 
\Phi_{\text{s}}^{(\alpha)}\right)\, i \chi_4^{(\alpha)}\chi_1^{(
\alpha)}\right] \quad .
\end{eqnarray}
\end{mathletters}
Here, the chiral boson fields $\Phi_{\text{c,s}}^{(\alpha)}$
represent fluctuations in the {\em total\/} charge and spin
density in the double-wire system; they are defined by
$\partial_x\Phi_{\text{c}}^{(\alpha)}/\pi=\varrho_{\uparrow\alpha
\text{U}}+\varrho_{\uparrow\alpha\text{L}}+\varrho_{\downarrow
\alpha\text{U}}+\varrho_{\downarrow\alpha\text{L}}$ and
$\partial_x\Phi_{\text{s}}^{(\alpha)}/\pi=\varrho_{\uparrow\alpha
\text{U}}+\varrho_{\uparrow\alpha\text{L}}-\varrho_{\downarrow
\alpha\text{U}}-\varrho_{\downarrow\alpha\text{L}}$,
respectively. The fields $\chi_j^{(\alpha)}=\chi_j^{(\alpha)
\dagger}$ are {\em Majorana\/} fermions. Their relation with the
original electronic degrees of freedom is highly nonlinear. (See
Appendix~\ref{bosref}.) Physical observables can be written in
terms of the bosonic normal modes as well as the Majorana
fermions. For example, the density of current flowing from the
upper wire to the lower one is given by $j=e/\hbar\cdot t\,
\sum_\alpha i\,\chi_1^{(\alpha)}\chi_3^{(\alpha)}$, and the
density of electrons with chirality $\alpha$ in the U (L) wire is
\begin{equation}\label{bosdens}
\tilde\varrho_{\alpha\text{U(L)}} = \sum_{\sigma} \varrho_{\sigma
\alpha\text{U(L)}}=\frac{\partial_x\Phi_{\text{c}}^{(\alpha)}}
{2\pi}\stackrel{-}{(+)} i\,\chi_2^{(\alpha)}\chi_3^{(\alpha)}
\quad .
\end{equation}

The advantage of representing $H$ in terms of the bosonic fields
$\Phi_{\text{c,s}}^{(\alpha)}$ and the Majorana fermions is that
interactions are absorbed into their respective
velocities.\cite{g2int} We find $v_{\text{c}}=(v_{\text{FU}}+
v_{\text{FL}}+[U_{\text{UU}}+U_{\text{LL}}+2 U_{\text{UL}}]/\pi
\hbar)/2$, $v_2=v_3=(v_{\text{FU}}+v_{\text{FL}}+[U_{\text{UU}}+
U_{\text{LL}}-2 U_{\text{UL}}]/\pi\hbar)/2$, and $v_{\text{s}}=v_1
=v_4=(v_{\text{FU}}+v_{\text{FL}})/2$. The parameter $\delta=e(
\nu_{\text{L}} V_{\text{L}}-\nu_{\text{U}} V_{\text{U}}) +
(v_{\text{FU}}+v_{\text{FL}})\kappa_{++}/2$ is given in terms of
the applied voltages and magnetic field, and
$H_{\text{corr}}^{(\alpha)}$ contains the effect of the wires
being not identical: $\Delta_{\text{c}}=(v_{\text{FU}}-
v_{\text{FL}} + [U_{\text{UU}} - U_{\text{LL}}]/\pi\hbar)/2$, and
$\Delta_{\text{s}}=(v_{\text{FU}}-v_{\text{FL}})/2$. For
comparison, it is useful to note the relations
\begin{mathletters}
\begin{eqnarray}
v_{\text{c}}&=&\frac{v_{\rho\text{U}}+v_{\rho\text{L}}}{2} +
\frac{U_{\text{UL}}}{\pi\hbar}\quad ,\\
v_{\text{s}}&=&v_1=v_4=\frac{v_{\sigma\text{U}}+
v_{\sigma\text{L}}}{2}\quad , \\
v_2&=&v_3=\frac{v_{\rho\text{U}}+v_{\rho\text{L}}}{2} -
\frac{U_{\text{UL}}}{\pi\hbar}\quad ,
\end{eqnarray}
\end{mathletters}
in terms of velocities defined in Sec.~\ref{pertsub}. When
inter-wire interactions are strong, i.e., $\pi\hbar|
\Delta_{\text{c}}/U_{\text{UL}}|\equiv r\ll 1$, we find
$v_{\text{c}}\approx v_{\rho+}$ and $v_2=v_3\approx v_{\rho-}$.

\subsection{Results for the density response function}
\label{ressec}

We consider the retarded real-time\cite{realnow} density response
function
\begin{equation}\label{retfunc}
{\mathcal{D}}_{\beta\beta^\prime}^{(\alpha)}(x,\tau) = -i\Theta
(\tau)\,\langle [\tilde\varrho_{\alpha\beta}(x, \tau), \tilde
\varrho_{\alpha\beta^\prime}(0,0)]\rangle
\end{equation}
for the special case of identical wires ($\Delta_{\text{c}}=
\Delta_{\text{s}}=0$) but with inter-wire interaction present.
This extends previous work\cite{fabrizio} to the experimentally
relevant situation in real double-wire systems. Hence, in the
following, we have $v_{\text{c}}\equiv v_{\rho+}$, $v_{\text{s}}
=v_1=v_4\equiv v_{\sigma}$, and $v_2=v_3\equiv v_{\rho-}$. The
density response function is then the sum ${\mathcal{D}}_{\beta
\beta^\prime}^{(\alpha)}={\mathcal{D}}_{\text{c}}^{(\alpha)}+
\eta_{\beta\beta^\prime}{\mathcal{D}}_{\text{n}}^{(\alpha)}$ of
a contribution from the total-charge mode,
${\mathcal{D}}_{\text{c}}^{(\alpha)}(x,\tau)=\Theta(\tau)\,
\delta^\prime\left(x - \alpha v_{\rho+}\tau\right)\,/2\pi$,
where the prime denotes differentiation with respect to the
argument of the delta function, and a term originating from the
Majorana fermions. Here $\eta_{\text{UU}}=\eta_{\text{LL}}=-
\eta_{\text{UL}}=-\eta_{\text{LU}}=1$. Further simplification
arises in the case $\delta=0$ (`on resonance' according to the
definition given in Ref.~\onlinecite{uz:01b}) where spatial
oscillations due to coherent motion of electrons between the two
wires can be expected to be largest. Details of our calculation
can be found in Appendix~\ref{respappend}. A characteristic length
scale emerges, given by $L_{\text{int}} = \pi\hbar|v_{\rho-}-
v_{\sigma}|/|2t|$, which measures the relative strength of
tunneling and interactions. Note that $v_{\rho-}$ differs from
$v_{\sigma}$ by the {\em difference\/} of intra and inter-wire
interactions. It approaches the spin-mode velocity not only in the
limit of weak interactions but also when interactions between and
within the wires are equal. On length scales that are large
compared to $L_{\text{int}}$, i.e., when $|x-\alpha\bar v\tau|>
L_{\text{int}}$, we find
\begin{eqnarray}
&&{\mathcal{D}}_{\text{n}}^{(\alpha)}(x,\tau)=\frac{\Theta(\tau)}
{2\pi}\frac{\bar v}{v_{\rho-}}\cos\left(\frac{2\pi x}{L_t}
\right)\nonumber \\ && \hspace{0.5cm} \times
\left[\bar v \delta(x - \alpha\bar v\tau)-v_{\rho-}\delta(x-
\alpha v_{\rho-}\tau)\right]/[(v_{\rho-}-\bar v)x] \, ,
\end{eqnarray}
showing oscillations with wave length $L_t=\pi\hbar\sqrt{v_{\rho-}
v_\sigma}/|t|$, and {\em two\/} propagation velocities $v_{\rho-}$
and $\bar v=2 v_{\rho-} v_\sigma/(v_{\rho-}+v_\sigma)$. On length
scales shorter than $L_{\text{int}}$, which can be of practical
relevance for strong asymmetric interactions, the neutral-mode
density response depends on $v_{\rho-}$ only:
${\mathcal{D}}_{\text{n}}^{(\alpha)}(x,\tau)=\Theta(\tau)\,
\delta^\prime\left(x - \alpha v_{\rho-}\tau\right)/2\pi$.

{\em Single-electron\/} Greens functions in tunnel-coupled 1D
electron systems have been shown\cite{fink:prb:93,fabrizio} to
exhibit {\em three\/} singularities as opposed to the two
associated with spin and charge degrees of freedom in the single
system. In one of the studies,\cite{fabrizio} the velocity of the
additional mode is given by the analog of $\bar v$. From our
results above, we see that this new mode also appears in the
density response and should therefore be observable in a
time-resolved experiment.\cite{obscav}

Let us briefly comment on the effect of deviations from the
ideal case considered here. When the wires are not identical,
$H_{\text{corr}}^{(\pm)}$ effectively introduce interactions
between bosonic normal modes and fictitious fermions. Additional
interaction terms appear when forward scattering between
left-movers and right-movers is included. In principle, these
could be treated within a self-consistent mean-field
approximation, yielding a Hamiltonian of the form $H-\sum_\alpha
H_{\text{corr}}^{(\alpha)}$ with renormalized parameters. From our
experience,\cite{uz:01b} we expect such nonidealities to suppress
the amplitude of coherent charge oscillations.

\section{Conclusions}

We have considered a model system for interacting tunnel-coupled
quantum wires where interactions between electrons of opposite
chirality are neglected. Signatures of spin-charge separation are
found to appear in the differential tunneling conductance. We give
explicit expressions for how these features depend on spin and
charge velocities in the wires as well as parameters measuring
charging of the wires due to the applied voltage.

Inclusion of nonchiral interactions which are present in real
quantum wires will not affect these predictions, as the locations
of these particular maxima in the DTC are captured correctly
already in the minimal (chiral) model. However, the values of
characteristic charge and spin velocities as well as the detailed
distribution of spectral weight will certainly be changed by the
presence of nonchiral interactions.\cite{balents} Most
importantly, additional shadow maxima can appear which we expect
to behave qualitatively similar to those present in the chiral
model.

Within a nonperturbative treatment of tunneling and interactions,
we find that spin-charge separation is manifested in the density
response function by the appearance of an additional mode which
could be observed in a time-resolved experiment. 

\acknowledgments

This work was supported by DFG Sonderforschungsbereich 195 and
the EU LSF programme. Useful discussions with D.~Boese, A.~Rosch,
M.~Sassetti, and A.~Yacoby are gratefully acknowledged.

\appendix

\section{Bosonization and refermionization formalism}
\label{bosref}

Here we provide details on how to rewrite the original interacting
Hamiltonian, given by Eqs.~(\ref{hamilton}), in terms of bosonic
and fictitious fermionic degrees of freedom to obtain the
equivalent but more easily tractable Hamiltonian displayed in
Eqs.~(\ref{rewrite}).
To keep notation simple, we consider here only the right-moving
part of the Hamiltonian $H$, denoted by $H_+$, which contains the
terms with $\alpha=+$ in Eqs.~(\ref{hamilton}). The remaining part
of $H$ describing left-movers is treated analogously. First we
apply the bosonization procedure outlined in
Ref.~\onlinecite{fink:prb:93}. Using the representation of
antibonding and bonding electron states in real space, given by
$\Psi_{\sigma,\pm}=(\psi_{\sigma+\text{L}}\pm\psi_{\sigma+
\text{U}})/\sqrt{2}$, we find $H_+=\sum_{\gamma=\text{c,s}}
H_\gamma^{(+)} + H_{\tilde{\text{c}}} + H_{\tilde{\text{s}}} +
\tilde{H}_{\delta}^{(+)} +\tilde{H}_{\text{corr}}^{(+)}$, with
$H_{\gamma}^{(+)}$ given by Eq.~(\ref{symmbos}), and 
\begin{mathletters}\label{intermed}
\begin{eqnarray}
H_{\tilde{\text{c}}}&=& \int_x \left\{\frac{\tilde v}{4\pi}\left(
\partial_x\Phi_{\tilde{\text{c}}}\right)^2 \right.\nonumber\\
&& \left. + \frac{\tilde g_4}{2}
\sum_\sigma\left( \Psi_{\sigma,-}^\dagger \Psi_{-\sigma,-}^\dagger
\Psi_{-\sigma,+}\Psi_{\sigma,+} + \text{H.c.}\right)\right\} \\
H_{\tilde{\text{s}}}&=& \int_x \left\{\frac{\tilde v}{4\pi}\left(
\partial_x\Phi_{\tilde{\text{s}}}\right)^2 \right.\nonumber\\
&& \left. + \frac{\tilde g_4}{2}
\sum_\sigma\left( \Psi_{\sigma,-}^\dagger \Psi_{-\sigma,+}^\dagger
\Psi_{-\sigma,-}\Psi_{\sigma,+} + \text{H.c.}\right)\right\} \\
\tilde{H}_{\delta}^{(+)}&=&-\frac{\delta}{2}\int_x\sum_\sigma
\left(\Psi_{\sigma,+}^\dagger \Psi_{\sigma,-} +
\Psi_{\sigma,-}^\dagger\Psi_{\sigma,+}\right) \\
\tilde{H}_{\text{corr}}^{(+)}&=& \pi \int_x \sum_{\sigma
\sigma^\prime}\left(\Delta v_{\text{F}}\,\delta_{\sigma
\sigma^\prime} + \Delta g_4\right)\nonumber \\ && \hspace{-1cm}
\times \left(\Psi_{\sigma,+}^\dagger\Psi_{\sigma,+}+\Psi_{\sigma,
-}^\dagger\Psi_{\sigma,-}\right)\left(\Psi_{\sigma^\prime,
+}^\dagger\Psi_{\sigma^\prime,-} + \text{H.c.}\right)
\end{eqnarray}
\end{mathletters}
Here the tunneling strength has been absorbed in different Fermi
wave vectors $k_{\text{F}\gamma}=\left(k_{\text{FU}}+k_{\text{FL}}
\right)/2 - \gamma\, t/\tilde v$ for the bonding and antibonding
fields $\Psi_{\sigma,\gamma}$. In addition to $\Phi_{\text{c,s}}$
that appeared previously, we have introduced new phase fields
defined via $\partial_x\Phi_{\tilde{\text{c}}}/\pi=\sum_\sigma
\left(\Psi_{\sigma,+}^\dagger\Psi_{\sigma,+}-\Psi_{\sigma,
-}^\dagger\Psi_{\sigma,-}\right)$ and $\partial_x
\Phi_{\tilde{\text{s}}}/\pi=\sum_\sigma\sigma\left(\Psi_{\sigma,
+}^\dagger\Psi_{\sigma,+}-\Psi_{\sigma,-}^\dagger\Psi_{\sigma,-}
\right)$. We have used the abbreviations $\tilde v =(v_{\text{FU}}
+v_{\text{FL}}+[U_{\text{UU}}+U_{\text{LL}}-2 U_{\text{UL}}]/2\pi
\hbar)/2$, $\tilde g_4=(U_{\text{UU}}+U_{\text{LL}}-2U_{\text{UL}}
)/4$, $\Delta v_{\text{F}} = (v_{\text{FU}}-v_{\text{FL}})/2$, and
$\Delta g_4 = (U_{\text{UU}}-U_{\text{LL}})/2$. A bosonization
identity\cite{vondelft} relates the fermionic operators
$\Psi_{\sigma,\gamma}$ to corresponding bosonic phase fields
$\phi_{\sigma,\gamma}=\left(\phi_{\text{c}}+\gamma\phi_{\tilde{
\text{c}}} +\sigma\left[\phi_{\text{s}}+\gamma\phi_{\tilde
{\text{s}}}\right]\right)/2$:
\begin{equation}\label{bosid}
\Psi_{\sigma,\gamma}(x)=\frac{1}{\sqrt{\mathcal N}}\, F_{\sigma,
\gamma}\, e^{i x k_{\text{F}\gamma}}\, e^{i\phi_{\sigma,\gamma}
(x)}\quad .
\end{equation}
Here, ${\mathcal N}$ is a normalization constant, while
$F_{\sigma,\gamma}$ denotes a Klein factor\cite{vondelft} which
acts as a ladder operator for particle species indexed by quantum
numbers $\sigma,\gamma$ and obeys fermionic commutation rules.
With the help of the bosonization identity, we can rewrite
products of Fermi operators appearing in Eqs.~(\ref{intermed})
entirely in terms of phase fields and Klein factors. For example,
we find
\begin{eqnarray}\label{firstbos}
&&\Psi_{\sigma,-}^\dagger \Psi_{-\sigma,-}^\dagger
\Psi_{-\sigma,+}\Psi_{\sigma,+}=\nonumber \\ &&\hspace{1cm}
{\mathcal N}^{-2}\, F_{\sigma,-}^\dagger F_{-\sigma,-}^\dagger
F_{-\sigma,+} F_{\sigma,+}\, e^{i\left(2\Phi_{\tilde{\text{c}}} -
4 t/\tilde v\right)}\quad .
\end{eqnarray}
This is how far the bosonization procedure was applied in
Ref.~\onlinecite{fink:prb:93} for the special case of identical
wires.

We now proceed to refermionize terms in the Hamiltonian containing
exponentials of phase fields. To this end, the phase fields
$\Phi_{\tilde{\text{c}},\tilde{\text{s}}}$ are used to define new
fermionic operators, essentially by applying the bosonization
identity in reverse:
\begin{equation}
\psi_{\tilde{\text{c}},\tilde{\text{s}}} = \frac{1}{\sqrt{\mathcal
N}}\, F_{\tilde{\text{c}},\tilde{\text{s}}}\, e^{i x
k_{\tilde{\text{c}},\tilde{\text{s}}}}\, e^{i\Phi_{\tilde
{\text{c}},\tilde{\text{s}}}(x)}\quad .
\end{equation}
Straightforward algebra shows that certain products of Klein
factors appearing in Eq.~(\ref{bosid}) exhibit the properties
required for Klein factors $F_{\tilde{\text{c}},\tilde{\text{s}}}
$. In particular, two equivalent representations can be found:
\begin{mathletters}\label{kleinf}
\begin{eqnarray}
F_{\tilde{\text{c}}}^{(1)} &=& F_{\uparrow,+}
F_{\downarrow, -}^\dagger \quad F_{\tilde{\text{s}}}^{(1)} =
F_{\downarrow,-} F_{\uparrow, -}^\dagger \quad ,\\
F_{\tilde{\text{c}}}^{(2)} &=& F_{\downarrow,+}
F_{\uparrow, -}^\dagger \quad F_{\tilde{\text{s}}}^{(2)} =
F_{\downarrow,+} F_{\uparrow, +}^\dagger \quad .
\end{eqnarray}
\end{mathletters}
As it turns out, products of Klein factors arising in bosonized
expressions for terms in Eqs~(\ref{intermed}) that are bilinear or
quadrilinear in fermion operators can be rewritten as products of
{\em two\/} Klein factors from the representations introduced in
Eqs.~(\ref{kleinf}). For example, we find
\begin{mathletters}
\begin{eqnarray}
\Psi_{\sigma,-}^\dagger \Psi_{-\sigma,-}^\dagger\Psi_{-\sigma,+}
\Psi_{\sigma,+}
&=& - F_{\tilde{\text{c}}}^{(1)}F_{\tilde{\text{c}}}^{(2)}\,\,
\frac{e^{i\left(2\Phi_{\tilde{\text{c}}} - 4 
t/\tilde v\right)}}{{\mathcal N}^2} \, , \\\label{specref1}
&=& i\,\psi_{\tilde{\text{c}}}\,\partial_x\, \psi_{\tilde{\text{c}
}}\quad .
\end{eqnarray}
\end{mathletters}
Here we applied the bosonization identity given in Eq.~(68) of
the first paper cited in Ref.~\onlinecite{naudpapers}. Similarly,
employing the identity $F_{\tilde{\text{s}}}^{(1)}
F_{\tilde{\text{c}}}^{(1)}=F_{\uparrow,-}^\dagger F_{\uparrow,+}$,
we obtain
\begin{mathletters}
\begin{eqnarray}
\Psi_{\uparrow,-}^\dagger \Psi_{\uparrow,+}&=& {\mathcal N}^{-1}
\, F_{\tilde{\text{s}}}^{(1)} F_{\tilde{\text{c}}}^{(1)} e^{i(
\Phi_{\tilde{\text{c}}}+\Phi_{\tilde{\text{s}}}-2 t/\tilde v)}
\,\, , \\ \label{specref2}
&=& \psi_{\tilde{\text{s}}}\, \psi_{\tilde{\text{c}}}\quad .
\end{eqnarray}
\end{mathletters}
Note that the {\em sign\/} of the refermionized terms given in
Eqs.~(\ref{specref1}) and (\ref{specref2}) is determined by the
particular arrangement of Klein factors in the initial bosonized
form as well as the correct representation of Klein factors
[Eqs.~(\ref{kleinf})] for the new fermions. All other terms in
Eqs.~(\ref{intermed}) that contain products of Fermi operators can
be treated analogously. As a result, we obtain finally
\begin{mathletters}
\begin{eqnarray}\label{fictcham}
H_{\tilde{\text{c}}}&=&\int_x\Big\{
\psi_{\tilde{\text{c}}}^\dagger\left(-i\tilde v\partial_x + 2 t
\right)\psi_{\tilde{\text{c}}}\nonumber \\ && -
\frac{\tilde{g}_4}{2\pi}\left[\psi_{\tilde{\text{c}}}^\dagger (-i
\partial_x )\psi_{\tilde{\text{c}}}^\dagger+\psi_{\tilde{\text{c}}
}(-i\partial_x )\psi_{\tilde{\text{c}}}\right]\Big\} \, ,\\
H_{\tilde{\text{s}}}&=&\int_x\Big\{
\psi_{\tilde{\text{s}}}^\dagger(-i\tilde v\partial_x)
\psi_{\tilde{\text{s}}} \nonumber \\ && + 
\frac{\tilde{g}_4}{2\pi}\left[\psi_{\tilde{\text{s}}}^\dagger (-i
\partial_x )\psi_{\tilde{\text{s}}}^\dagger+\psi_{\tilde{\text{s}}
}(-i\partial_x )\psi_{\tilde{\text{s}}}\right]\Big\} \, , \\
\tilde H_\delta^{(+)} &=& \frac{\delta}{2}\int_x\left(
\psi_{\tilde{\text{c}}}-\psi_{\tilde{\text{c}}}^\dagger\right)
\left(\psi_{\tilde{\text{s}}}+\psi_{\tilde{\text{s}}}^\dagger
\right) \, , \\
\tilde H_{\text{corr}}^{(+)} &=&\Delta_{\text{c}}\int_x \,
\partial_x \Phi_{\text{c}}\,\left(\psi_{\tilde{\text{c}}}^\dagger
-\psi_{\tilde{\text{c}}}\right)\,\left(
\psi_{\tilde{\text{s}}}^\dagger + \psi_{\tilde{\text{s}}}\right)
\nonumber \\ && + \Delta_{\text{s}}\int_x \partial_x 
\Phi_{\text{s}}\,\left(\psi_{\tilde{\text{c}}}^\dagger
+\psi_{\tilde{\text{c}}}\right)\,\left(
\psi_{\tilde{\text{s}}}^\dagger - \psi_{\tilde{\text{s}}}\right)
\, .
\end{eqnarray}
\end{mathletters}
Using the representation of chiral Dirac fermions in terms of
Majorana fermions according to $\psi_{\tilde{\text{c}}}=
(\chi_1+i\chi_2)/\sqrt{2}$ and $\psi_{\tilde{\text{s}}}=(\chi_3+i
\chi_4)/\sqrt{2}$, we obtain Eqs.~(\ref{rewrite}). We would like
to stress the point that only the correct treatment of Klein
factors enables the unambiguous determination of the velocities of
the Majorana modes. Our approach differs from the one taken in
Ref.~\onlinecite{fabrizio} where fictitious fermions were defined
via phase fields that are linear combinations of charge and
spin-density phase fields of the {\em individual\/} quantum
wires and not those of the bonding and antibonding states used
here and in Ref.~\onlinecite{fink:prb:93}.

\section{Calculation of the density response function}
\label{respappend}

We obtain the retarded real-time\cite{realnow} density response
function defined in Eq.~(\ref{retfunc}) by generalizing methods
described in Refs.~\onlinecite{naudpapers} to our case of
interest. Again, to avoid unnecessary repetition, we focus only on
the case of right-movers, i.e., $\alpha=+$. We start by
considering the time-ordered Greens function
\begin{equation}
{\mathcal{D}}_{\beta\beta^\prime}^T(x,\tau) = -i\langle
T \tilde\varrho_{\alpha\beta}(x, \tau) \tilde\varrho_{\alpha
\beta^\prime}(0,0)]\rangle\quad .
\end{equation} 
Applying the representation of charge densities in terms of the
free boson and fictitious-fermion degrees of freedom given in
Eq.~(\ref{bosdens}) and specializing to the case
$\delta=\Delta_{\text{c}}=\Delta_{\text{s}}=0$ yields with
$\eta_{\text{UU}}=\eta_{\text{LL}}=-\eta_{\text{UL}}=-
\eta_{\text{LU}}=1$
\begin{mathletters}
\begin{eqnarray}
{\mathcal{D}}_{\beta\beta^\prime}^T&=&{\mathcal{D}}_{\text{c}}^T +
\eta_{\beta\beta^\prime}\,{\mathcal{D}}_2^T {\mathcal{D}}_3^T
\quad , \\
{\mathcal{D}}_{\text{c}}^T(x,\tau) &=& \frac{-i}{4\pi^2}\, \langle
T \partial_x\Phi_{\text{c}}(x,\tau)\partial_x\Phi_{\text{c}}(0,0)
\rangle \, , \\
{\mathcal{D}}_2^T(x,\tau) &=& \langle T\, i\chi_2(x,\tau)\,i
\chi_2(0,0)\rangle\quad , \\
{\mathcal{D}}_3^T(x,\tau) &=& -i \langle T \chi_3(x,\tau)
\chi_3(0,0)\rangle\quad .
\end{eqnarray}
\end{mathletters}
For the special case considered here ($\delta=\Delta_{\text{c}}=
\Delta_{\text{s}}=0$), the correlation functions
${\mathcal{D}}_{\text{c}}^T$ and ${\mathcal{D}}_3^T(x,\tau)$ are
trivial:
\begin{mathletters}
\begin{eqnarray}
{\mathcal{D}}_{\text{c}}^T(x,\tau) &=& \frac{i}{4\pi^2}\,
\left[x-v_{\text{c}}\tau + i \epsilon\,{\mathrm{sgn}}(\tau)
\right]^{-2}\, , \\
{\mathcal{D}}_3^T(x,\tau)&=&\frac{1}{2\pi}\,\left[x-v_3\tau + i
\epsilon\, {\mathrm{sgn}}(\tau)\right]^{-1}\, .
\end{eqnarray}
\end{mathletters}
Calculation of ${\mathcal{D}}_2^T$ is nontrivial, as $\chi_2$ is
coupled to $\chi_1$ through the term $H_t^{(+)}$. It is therefore
advantageous to work in the representation of the fictitious Dirac
fermion field $\psi_{\tilde{\text{c}}}$ whose Hamiltonian is given
by Eq.~(\ref{fictcham}), and to use the identity
\widetext
\top{-2.8cm}
\begin{equation}\label{long}
{\mathcal{D}}_2^T(x,\tau) = \frac{1}{2}\left\{\langle T\,
\psi_{\tilde{\text{c}}}(x,\tau)\psi_{\tilde{\text{c}}}(0,0)\rangle
+ \langle T\,\psi_{\tilde{\text{c}}}^\dagger(x,\tau)\psi_{\tilde
{\text{c}}}^\dagger(0,0)\rangle - \langle T\,
\psi_{\tilde{\text{c}}}^\dagger(x,\tau)\psi_{\tilde{\text{c}}}(0,
0)\rangle-\langle T\,\psi_{\tilde{\text{c}}}(x,\tau)\psi_{\tilde
{\text{c}}}^\dagger(0,0)\rangle\right\}\quad .
\end{equation}
The correlation functions for the field $\psi_{\tilde{\text{c}}}$
appearing in Eq.~(\ref{long}) can be calculated.\cite{naudpapers}
After some algebra, we obtain
\begin{equation}\label{longres}
{\mathcal{D}}_2^T(x,\tau) = \frac{-i}{4\pi}\frac{v_1}{\tilde
g_4} \frac{\cos\left(2\pi\frac{x}{L_t}\right)}{x} -{\mathrm{sgn}}
(\tau)\frac{x - v_1\tau + i \epsilon\,{\mathrm{sgn}}(\tau)}{2
\tilde g_4 x}\,\int_0^\infty \frac{d \omega}{2\pi} \, e^{i\omega
\tilde\tau}\,\cos[\kappa(\omega)\, X] \quad .$$
\end{equation}
\bottom{-2.7cm}
\narrowtext
\noindent
Here we used the notation\cite{naudpapers} $X=x/v_1 v_2$, $\tilde
\tau = {\mathrm{sgn}}(\tau)[\tilde v X - \tau] + i\epsilon$, and
$\kappa(\omega) = \sqrt{4 v_1 v_2 t^2 + \tilde g_4^2\omega^2}$.
The $\omega$-dependence of $\kappa(\omega)$ introduces the
crossover scale $L_{\text{int}}=\pi\hbar(v_2-v_1)/2t$ that was
found previously\cite{fabrizio} within a different approach.
Inspecting our general result~(\ref{longres}) for
${\mathcal{D}}_2^T(x,\tau)$ in the limit of length scales that are
larger or smaller than $L_{\text{int}}$ and transforming to the
retarded correlation function yields the results quoted in
Sec.~\ref{ressec}.

\widetext 


\begin{thebibliography}{10}

\bibitem{voit:reprog:94}
For a recent review, see J.~Voit,
\newblock Rep. Prog. Phys. {\bf 57}, 977 (1994).

\bibitem{fdmh:jpc:81}
F.~D.~M. Haldane,
\newblock J. Phys. C {\bf 14}, 2585 (1981).

\bibitem{velocaveat}
At small wave vectors, the dispersion of the phonon modes is
linear for screened Coulomb interactions that we expect to be
present in real quasi-1D samples.

\bibitem{mahan}
G.~D. Mahan,
\newblock {\em Many-Particle Physics} (Plenum Press, New York,
1990).

\bibitem{med:prb:92}
V.~Meden and K.~Sch{\"o}nhammer,
\newblock Phys. Rev. B {\bf 46}, 15753 (1992).

\bibitem{voit:prb:93}
J.~Voit,
\newblock Phys. Rev. B {\bf 47}, 6740 (1993).

\bibitem{powerlaw}
In addition to spin-charge separation, a Luttinger liquid can
exhibit power-law behavior of electronic correlation functions.
As will become clear below, we focus here entirely on effects due
to spin-charge separation, using a model for a Luttinger liquid
exhibiting no anomalous correlation-function exponents.

\bibitem{photo}
B. Dardel, D. Malterre, M. Grioni, P. Weibel, Y. Baer, and 
F. L\'evy,
\newblock Phys. Rev. Lett. {\bf 67}, 3144 (1991); 
C. Kim, A.~Y. Matsuura, Z.-X. Shen, N. Motoyama, H. Eisaki, S.
Uchida, T. Tohyama, and S. Maekawa,
\newblock Phys. Rev. Lett. {\bf 77}, 4054 (1996);
D. Orgad, S.~A. Kivelson, E.~W. Carlson, V.~J. Emery, X.~J. Zhou,
and Z.~X. Shen,
\newblock Phys. Rev. Lett. {\bf 86}, 4362 (2001).

\bibitem{alt:prl:99}
A.~Altland, C.~H.~W. Barnes, F.~W.~L. Hekking, and A.~J.
Schofield, \newblock Phys. Rev. Lett. {\bf 83}, 1203 (1999)

\bibitem{balents}
D. Carpentier, C. Pe{\c c}a, and L. Balents, cond-mat/0103193.

\bibitem{amir}
O.~M. Auslaender, A. Yacoby, R. de~Picciotto, K.~W. Baldwin, L.~N.
Pfeiffer, and K. West, Science {\bf 295}, 825 (2002).

\bibitem{before}
Momentum-resolved tunneling between low-dimensional electron
systems has been used before to measure single-electron
properties. For 2D--to--2D tunneling, see, e.g., J.~P.
Eisenstein, T.~J. Gramila, L.~N. Pfeiffer, and K.~W. West,
\newblock Phys. Rev. B {\bf 44}, 6511 (1991); L.~Zheng and A.~H.
MacDonald, \newblock Phys. Rev. B {\bf 47}, 10619 (1993). Studies
of 1D--to--2D tunneling can be found in B.~Kardyna\l , C.~H.~W.
Barnes, E.~H. Linfield, D.~A. Ritchie, J.~T. Nicholls, K.~M.
Brown, G.~A.~C. Jones, and M. Pepper,
\newblock Phys. Rev. B {\bf 55}, R1966 (1997);
Ref.~\onlinecite{alt:prl:99}; M.~Governale, M.~Grifoni, and
G.~Sch{\"o}n, \newblock Phys. Rev. B {\bf 62}, 15996 (2000).

\bibitem{uz:01b}
D.~Boese, M.~Governale, A.~Rosch, and U.~Z\"ulicke, Phys. Rev. B
{\bf 64}, 085315 (2001).

\bibitem{llnores}
In a generic Luttinger liquid, these resonances will be suppressed
due to vanishing spectral weight at the Fermi energy.

\bibitem{buttbefore}
M. B\"uttiker, H. Thomas, and A. Pr\^{e}tre, Phys. Lett. A {\bf
180}, 364 (1993); M. B\"uttiker, J. Phys.: Condens. Matter {\bf
5}, 9361 (1993); M. B\"uttiker and T. Christen, in: {\it Mesoscopic
Electron Transport}, edited by L.~L. Sohn, L.~P. Kouwenhoven, and
G. Sch\"on (Kluwer Academic, Dordrecht, 1997), pp.\ 259-289.

\bibitem{balentscomm}
Related work\cite{balents} has considered tunneling close to a
resonance point with $\alpha=-\alpha^\prime$, neglecting any
charging effects.

\bibitem{addmaxcomm}
In addition to the four maxima exhibited in the chiral model,
typically less pronounced shadow maxima can appear when non-chiral
interactions are present. See, e.g., Ref.~\onlinecite{balents}.

\bibitem{pertscale}
See, e.g., E. Arrigoni, Phys. Rev. Lett. {\bf 83}, 128 (1999), and
Ref.~\onlinecite{balents}. For the model of a Luttinger liquid
considered here, exhibiting spin-charge separation but no
anomalous power laws, the energy scale below which perturbation
theory breaks down is given, as in the noninteracting limit, by
the tunneling strength $|t|$.

\bibitem{dircoupl}
J.~A. {del Alamo} and C.~C. Eugster, Appl. Phys. Lett. {\bf 56},
 78  (1990).

\bibitem{fink:prb:93}
A.~M. Finkel'stein and A.~I. Larkin,
\newblock Phys. Rev. B {\bf 47}, 10461 (1993).

\bibitem{fabrizio}
M. Fabrizio and A. Parola, Phys. Rev. Lett. {\bf 70}, 226 (1993);
M. Fabrizio, Phys. Rev. B {\bf 48}, 15838 (1993).

\bibitem{lin:prb:98}
H. Lin, L. Balents, and M.~P.~A. Fisher,
\newblock Phys. Rev. B {\bf 58}, 1794 (1998).

\bibitem{naudpapers}
J.~D. Naud, L.~P. Pryadko, and S.~L. Sondhi,
\newblock Nucl. Phys. B {\bf 565}, 572 (2000);
\newblock Phys. Rev. B {\bf 63}, 115301 (2001).

\bibitem{Zeeman}
Here we neglect Zeeman splitting which is typically small for
magnetic fields needed to tune into the resonance point with
$\alpha=\alpha^\prime=+$. It can be included straightforwardly,
and one of its effects is a doubling of features in the DTC. See,
e.g., S. Rabello and Q. Si, cond-mat/0008065, and
Ref.~\onlinecite{balents}.

\bibitem{weakt}
We assume that tunneling is not strong enough to require a
self-consistent treatment of charging effects in the wires.

\bibitem{naudcomment}
Tunnel-coupled edge channels in quantum-Hall bilayers with each
layer having a {\em fractional\/} filling factor $\le 1$ have been
considered in Ref.~\onlinecite{naudpapers}.

\bibitem{LLcharge}
Y.~M. Blanter, F.~W.~J. Hekking, and M. B\"uttiker, Phys. Rev.
Lett. {\bf 81}, 1925  (1998); R. Egger and H. Grabert, Phys. Rev.
B {\bf 55},  9929  (1997), {\it ibid.} {\bf 58}, 13275(E) (1998).

\bibitem{replacecav}
The validity of Eqs.~(\ref{slopes}) was established by direct
comparison with our numerical results.

\bibitem{vondelft}
J. von~Delft and H. Schoeller, Ann. Phys. (Leipzig) {\bf 7}, 225
(1999); and references therein.

\bibitem{levitov}
L.~S. Levitov and A.~V. Shytov, cond-mat/9510006 (unpublished).

\bibitem{g2int}
This is true within our chiral model where interactions between
right-movers and left-movers are excluded. Taking them into
account introduces interaction terms of the kind $\chi_j^{(\alpha)
}\chi_k^{(\alpha)}\chi_j^{(-\alpha)}\chi_k^{(-\alpha)}$ that can
be treated, e.g., using mean-field theory.

\bibitem{realnow}
The reader be advised of our double use of the symbol $\tau$ as an
imaginary time in Sec.~\ref{pertsub} and a real time in
Sec.~\ref{ressec} and Appendix~\ref{respappend}.

\bibitem{obscav}
The difference $\bar v - v_{\sigma}=(v_{\rho-}-v_{\sigma})\frac
{v_\sigma}{(v_{\rho-}+v_{\sigma})}$ should be large enough to
enable observation in realistic systems\cite{amir} where
$v_{\rho-}$ is expected to be bigger but of the same order as
$v_{\sigma}$.

\end{thebibliography}
\end{document}